\documentclass[10pt,conference]{IEEEtran}

\usepackage{amsmath}
\usepackage{url}
\usepackage[pdftex]{graphicx}
\usepackage[utf8x]{inputenc}
\usepackage[warn]{textcomp}
\usepackage{graphicx}
\usepackage{amsfonts}
\usepackage{bm}
\usepackage{statmath}
\usepackage{mathtools,amssymb}
\usepackage{setspace}

\renewcommand{\IEEEkeywordsname}{Keywords}

\begin{document}

\title{Selectivity Gain in Olfactory Receptor Neuron at Optimal Odor Concentration}

\author{\IEEEauthorblockN{Alexander Vidybida}
\IEEEauthorblockA{Bogolyubov Institute for Theoretical Physics of the National Academy of Sciences of Ukraine, Kyiv, Ukraine 03143\\
Email: vidybidaok@gmail.com, \url{http://vidybida.kiev.ua}}
}

\maketitle

\begin{abstract}
Recently, it has been discovered that for the selectivity gain due to fluctuations
in the olfactory receptor neuron (ORN)
there exists the optimal concentration of odors at which increased selectivity
manifested most. We check what could be the gain value 
at that concentration by modeling ORN as a leaky integrate-and-fire neuron
with membrane
populated by receptor proteins (\textbf{\textit{R}}) which bind and release odor molecules randomly.
Each bound (\textbf{\textit{R}}) opens a depolarizing channel, Eq. (\ref{ORNm}), below.
\end{abstract}

\begin{IEEEkeywords}
 ORN, selectivity, receptor proteins, fluctuations
\end{IEEEkeywords}



\section{Introduction}

It is known from sensory biology that selectivity to stimulus of any modality
grows up as the response evoked by the stimulus propagates from primary sensory receptors
to more central brain areas, see e.g. \cite{Rust2010} for vision.
Better selectivity of projection neurons (stage 3 in Table \ref{Tab1})
as compared to ORNs (stage 2 in Table \ref{Tab1}) was
reported in \cite{Duchamp1982}. In parallel,
K. Persaud and G. Dodd \cite{Persaud1982} formulated a general principle for constructing
a selective electronic nose from poor selective primary units utilizing
morphology similar to that of biological primary olfactory pathways,
and proposed the combinatorial code for better odors discrimination.

Normally, the selectivity of
receptor proteins (stage 1 in Table \ref{Tab1}) is considered as
indistinguishable from that of corresponding ORN, e.g. \cite{Galizia2010}. 
This might not be the case if odors are applied in low concentrations,
 when fluctuations of receptors number bound with odor become essential.
It was shown theoretically for a rudimentary model of ORN, \cite{Vidybida2022}, that ORN's selectivity
in this case can possibly be higher than the that of $R$.
In this note, we check this possibility for the leaky integrate-and-fire neuronal model
with conductivity which fluctuates due to random binding-releasing of odor molecules by $R$
at the optimal (Sec. \ref{OC}) odors concentration.
{\small 
\begin{table}[h]
\caption{\label{Tab1}Selectivity build up steps in a biological olfactory system
(modified from \cite{Vidybida2022})}
\centering
\begin{tabular}{l |c | c}
& {\bf constructive element} & {\bf measure of response} \\
\hline
&\\
1. & receptor proteins & fraction of bound receptors     \\
&$\displaystyle \downarrow$ &   $\displaystyle \downarrow$ \\
2. & receptor neurons & {mean firing rate}\\
&$\displaystyle \downarrow$ &  $\displaystyle \downarrow$  \\
3. & projection neurons & {mean firing rate}\\
&   and antennal lobe     & and combinatorial code\\
&$\displaystyle \downarrow$ &  $\displaystyle \downarrow$  \\
4. & olfactory cortex & spatio-temporal activity in \\
&        & higher cortical circuits\\
\hline
\end{tabular}\medskip
\end{table} 
}

\section{Methods}
\subsection{Model of ORN}\label{ORN}

As model ORN we use the leaky integrate-and-fire model with fluctuating
conductance input, similar to that used for another purpose in \cite{Kuhn2004}
(see also \cite{Vidybida1996}):
\begin{equation}\label{ORNm} 
c_M\frac{dV(t)}{dt}=-g_l(V(t)-V_{rest}) - n(t)g_{R}(V(t)-V_{e}),
\end{equation}
where $V(t)$ --- is the membrane voltage; $V_{rest}$ --- is the resting voltage;
$c_M$ --- is the total capacity of ORN's membrane;
$g_l$ --- is the total leakage through it; 
$V_{e}$ --- is the reversal potential for current through open $R$;
$n(t)$ --- is the fluctuating number of open channels at moment $t$ due 
to odor molecules bound with receptors $R$;
$g_{R}$ --- is the conductance of a single open channel. 
The total number of $R$ in the ORN is denoted by $N$.
Here we adopt the paradigm: ``one bound $R$ $\to$
one open channel'', which is characteristic for insects, \cite{Sato2008}.
The model given by  (\ref{ORNm}) is extended with the triggering threshold
$V_{th}$: if the voltage $V(t)$ becomes equal to $V_{th}$ the ORN fires an
output spike and appears in its resting state with $V(t)=V_{rest}$.

In the case of $n(t)$ and $V(t)$ not depending of $t$ 
(no fluctuations are taken into account 
and all transients are completed) the number $N_0$ of $R$ bound with odor
required to trigger ORN can be found from (\ref{ORNm}):
\begin{equation}\label{N0}
N_0 = (g_l(V_{th}-V_{rest}))/(g_{R}(V_e-V_{th})).
\end{equation}
Our purpose is to consider concentrations at which the mean over time number $\overline{n}$
 of bound $R$ is close to $N_0$ and introduce fluctuations into (\ref{ORNm}).
For this regime of odor perception, selectivity of ORN appears to be considerably better than 
that of its receptors $R$.

\subsection{Selectivity of receptor proteins}

We describe interaction between receptor protein $R$ and an analyte $A$ by the following
association-dissociation reaction:
\begin{equation}\label{chr}
A+R\quad{\Large \mathrel{\substack{k_+\\\rightleftharpoons\\ k_-}}}\quad AR\,.
\end{equation}
If another analyte $A'$ is presented to a set of $R$ at the same concentration $[A]=c$ then
similar association-dissociation reaction   takes place with different rate
constants $k'_+$, $k'_-$. From the chemical point of view, the $R$ is able
to discriminate between $A$ and $A'$ if corresponding dissociation constants
$K$, $K'$ differ, where
\begin{equation}\label{K}
K=[A][R]/[AR] \quad\text{(at equilibrium.)}
\end{equation}
%
Supposing $K>K'$, chemical selectivity $h$ of $R$ might be expressed in terms of dissociation constants
as follows:
\begin{equation}\label{h}
h = (K-K')/K\,.
\end{equation}
But, neither $R$, nor the whole neuron have knowledge of $[A]$
required to determine $K$. Therefore, expressing selectivity 
through $K$ at this the very first stage of
odor perception seems being not relevant to how a nose operates in the field. 
Instead of $K$, similar to \cite{Vidybida2022} we use
the probability $p$ that a receptor $R$ is bound with $A$,
that at the same time is the fraction of  bound $R$ :
\begin{equation}\label{p}
p = [AR]/([AR]+[R]).
\end{equation}
The probability $p$ characterizes the initial response to analyte
in the set of $R$ belonging to a single ORN. 
As selectivity of $R$ we mean the selectivity 
of this initial response. Namely, if with analyte $A'$ we have
$p'<p$ then we define selectivity of $R$ with respect to $A$, $A'$
as
\begin{equation}\label{SR}
S_R = (p-p')/p\,.
\end{equation}

\subsection{Selectivity of ORN}
Denote as $F$, $F'$ the mean firing rate of ORN if analytes $A$, $A'$ are 
separately applied in the same concentration. 
As ORN's selectivity $S_{ORN}$ we take the following quotient:
\begin{equation}\label{SORN}
S_{ORN}=(F_1-F_2)\,/\,F_1\,.
\end{equation}

\subsection{Selectivity gain}

We assume here that the concentration $c$ ensures that the mean number of bound 
receptors $pN$, $p'N$ is close to the firing threshold $N_0$. In this case, the instantaneous
number $n(t)$ will cross the firing threshold $N_0$ randomly due to thermal fluctuations 
both for $A$ and $A'$. Fluctuations of this type can be observed experimentally, e.g. \cite{Menini1995}.
The rates $F$, $F'$ will be heavily dependent on those fluctuations.
We expect that the selectivity of ORN in this regime will be better then that of $R$
similarly as it was shown in \cite{Vidybida2022} for a simpler model.

In order to compare selectivity of ORN with that of its
receptor proteins we define the selectivity gain $g$
as follows:
\begin{equation}\label{g}
g=S_{ORN}\,/\,S_R\,.
\end{equation}

\subsection{Simulation algorithm outline}

\begin{figure}[t]
\centering
\includegraphics[width=0.16\textwidth, angle=-90]{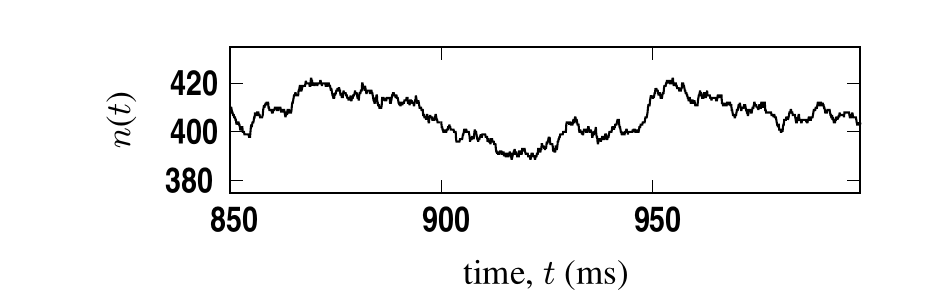}
\caption{\label{ntlong} Realization of stochastic process $n(t)$ near the end of long sniff.
 }
\end{figure}

\begin{figure}[b]
\includegraphics[width=0.16\textwidth, angle=-90]{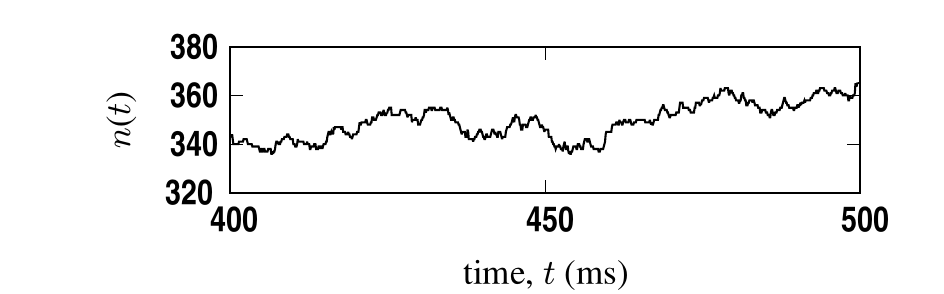}
\caption{\label{ntshort} Realization of stochastic process $n(t)$ near the end of short sniff.
 }
\end{figure}

We solve (\ref{ORNm}) numerically with the time step $dt=0.1$ ms.
The stochastic process $n_k\equiv n(kdt)$ is described as a Markov chain with
 transition matrix:
\begin{equation}\label{TM}
p(j\mid i),\quad i=0,\dots, N,\quad j=0,\dots,N.
\end{equation}
In our case,
 $p(j\mid i)$ gives the probability to have $n_{k+1}=j$ bound receptors at the moment $(k+1)dt$,
 provided that at the moment $kdt$ there were $n_k=i$ bound receptors (for any $k=0,1,\dots$).
The transition matrix is calculated in advance based on the concrete values for
$dt$, $c$, $k_+$, $k_-, N$ for both analytes. 
See Figs. \ref{ntlong}, \ref{ntshort} for examples of $n(t)$ realization.
The simulation of both deterministic electric transients and stochastic process
$n_k,~k=1,2,\dots$ in (\ref{ORNm})  is running in parallel: Having $n_k$ 
and $V(kdt)$ we firstly calculate the $V(kdt+dt)$ and then calculate the next random value
$n_{k+1}$ with the help of (\ref{TM}). It is decided that there was a spike
between $kdt$ and $(k+1)dt$ if $V(kdt+dt)\ge V_{th}$. In this case, 
the calculated  $V(kdt+dt)$
is replaced with $V_{rest}$, but $n_{k+1}$ remains untouched.

\section{Results}

The $R$ and $A$ parameters are chosen as follows:\smallskip

\begin{center}
\begin{tabular}{ccc}
$g_R$ (nS)    &  $k_+$ (ms$^{-1}$$\,$M$^{-1}$)   & $k_-$ (ms$^{-1}$)\\
\hline\\[-1em]
0.015       &  209                     &  7.90$\cdot 10^{-3}$
\end{tabular}
\end{center}
\smallskip

For the less affine odor $A'$ we assume the same $k_+$ and bigger 
releasing rate: $k'_- > k_-$. A concrete value for $k'_-$ is calculated based
on the assumed receptor selectivity $S_R$ with the help of  (\ref{p}),
(\ref{SR}). 
(The program gets as input the $S_R$ assumed value, see Tables \ref{long}, \ref{short}.)
E.g., if assumed $S_R=0.01$ then $k'_-=7.98\cdot 10^{-3}$ ms$^{-1}$
and smaller for smaller  $S_R$.

Electric parameters used in (\ref{ORNm}) are  as follows:\smallskip

\begin{center}
\begin{tabular}{ccccc}
$c_M$ (pF) & $g_l$ (nS) & $V_{rest}$ (mV)    &  $V_{th}$ (mV)   & $V_e$ (mV)\\
\hline\\[-1em]
4.26 & 0.213 & -80       &  -54                     &  -53
\end{tabular}
\end{center}\smallskip

Now, (\ref{N0}) gives $N_0=369$. For the total number of receptors
we choose $N=2556000$. These parameters are chosen based mainly on the paper
\cite{Kaissling2001}. As regards $V_e$, experimental data was not available.
The only thing which is certain is that $V_e>V_{th}$, otherwise spiking due to odor
would be impossible.

\subsection{Optimal concentration}\label{OC}
For a more rudimentary model considered in \cite{Vidybida2022} it was found in
\cite{Vidybida2023} that the selectivity gain due to fluctuations is better 
pronounced if concentration of any odor is taken in the vicinity of $c_0$, where
\begin{equation}\label{c0}
c_0=K(N_0-1)/(N-N_0).
\end{equation}
In the model of \cite{Vidybida2022}, achieving the value $N_0$ by $n(t)$ results in
immediate firing. This is not the case for the model (\ref{ORNm}), (\ref{N0}), above,
due to the relaxation processes in the membrane. But not having another estimate
for $c_0$, we take the one given by (\ref{c0}) with $N_0$ given by (\ref{N0}),
which for above given parameters and $K=k_-/k_+$
is $c_0=5.44\cdot10^{-09}$ M. We choose $c=c_0$ for
all numeric simulations.

\subsection{Long sniff paradigm}
The ORN starts at its resting state with no bound $R$. The number $n(t)$
relaxes from zero to its mean number with relaxation time $\tau = 127$ ms,
Fig. \ref{ntlong}.
The sniff / run duration is 1 sec and 1000 runs have been performed
without resetting the random numbers generator (the {\tt knuthran2002}
generator from GNU Scientific Library, \cite{Galassi2009}; 
three different seeds were used).
This is equivalent to have a single sniff with 1000 ORNs converging onto
a single glomerulus. The obtained mean firing rate per neuron
in a single sniff is 55 Hz for $A$ 
and 45 Hz for $A'$ if $S_R=0.01$ was chosen.
The selectivity gain depends on the chosen $S_R$ as 
shown in the Table \ref{long}.
\begin{table}[t]
\caption{\label{long}
Selectivity gain for long sniffs}
\centering
\begin{tabular}{rc|ccccc}
input: & $S_R$ & 0.1 & 0.01 & 0.001 & $10^{-4}$ & $10^{-5}$\\
\hline
\rule{0pt}{12pt}
output:& $S_{ORN}$ &  0.96  &  0.18  &  0.024  &  0.007  &   0.005      \\          
\hline
\rule{0pt}{12pt}
output:& $g$       & 9.6   &  18  &  23.5  &  72  &    538      \\
\end{tabular}
\end{table}

\subsection{Short sniff paradigm}
Here, sniff duration is 500 ms, and  5000 sniffs was performed. This is equivalent to
5000 identical ORNs converging onto a single glomerulus and performing a single sniff
500 ms long. One example of $n(t)$ relaxing to its mean value is given in 
the Fig. \ref{ntshort}.
The results obtained are shown in the Table \ref{short}.
\begin{table}[b]
\caption{\label{short}
Selectivity gain for short sniffs}
\centering
\begin{tabular}{rc|ccccc}
input: & $S_R$ & 0.1 & 0.01 & 0.001 & $10^{-4}$ & $10^{-5}$\\
\hline
\rule{0pt}{12pt}
output:& $S_{ORN}$ &  0.97  &  0.23  &  0.042  &  0.021  &   0.019     \\          
\hline
\rule{0pt}{12pt}
output:& $g$       & 9.7   &  23  &  41.7  &  208  &    1867     \\
\end{tabular}
\end{table}

\section{Conclusions and discussion}
In this note, we used numerical simulation in order to compare selectivity
of ORN with that of its receptor proteins ($R$). As the neuronal model we use
the leaky integrate-and-fire one with fluctuating conductivity due to 
random nature of odor binding-releasing by  receptors.  The possible selectivity gain,
see Tables \ref{long}, \ref{short}, appears to be large provided that
odors are applied in the optimal concentration (\ref{c0}) and $R$ has a poor selectivity.
The limitation to have the fixed concentration can be alleviated in ORN by known biophysical
mechanisms able to change effective concentration, \cite{Nagashima2010}, and   
the threshold $N_0$, \cite{Bryche2021}. 
In artificial bio-inspired sensors, \cite{Hurot2020}, there should be wider technical possibilities
to ensure that odors are sensed by ORNs in the optimal for selectivity gain concentration.\medskip

{
\footnotesize
\begin{spacing}{0.9}
Acknowledgment. This work was sponsored in the framework of the Fundamental research program of the Branch of Physics and Astronomy of the National Academy of Sciences of Ukraine (project № 0120U101347 “Noise-induced dynamics and correlations in non-equilibrium systems”) and by the private American Simons Foundation.
\end{spacing}
}

\vspace*{0.05cm}

\end{document}